# Route Extrapolation for Source and Destination Camouflage in Wireless Ad Hoc Networks

M. Razvi Doomun
Department of Computer Science and Engineering
University of Mauritius
Reduit, Mauritius
r.doomun@uom.ac.mu

K.M. Sunjiv Soyjaudah
Faculty of Engineering
University of Mauritius
Reduit, Mauritius
ssoyjaudah@uom.ac.mu

*Abstract*— In wireless ad hoc networks, protecting source and destination nodes' location privacy is a challenging task due to malicious traffic analysis and privacy attacks. Existing solutions, such as incorporating fake source-destination pairs in the network, provide some privacy of real source and destination nodes against attackers. Moreover, ad hoc networks need stronger privacy protection against powerful global attacker which has knowledge of overall network topology and, that can also eavesdrop and visualize network-wide data transmissions. In this paper, we present a novel privacy technique, EXTROUT: Route Extrapolation to camouflage the real source and destination nodes along an extended path in an ad hoc network. We demonstrate that the privacy level achieved with EXTROUT is higher and more effective against a global attacker, when compared to fake source-destination nodes privacy scheme.

## I. INTRODUCTION

In wireless ad hoc networks, source and destination nodes communicate directly if they are within transmission range of each other, else they rely on intermediate nodes for packet routing. Broadcast transmission in ad hoc wireless communication is prone to malicious traffic analysis attacks [1][2]. In general, traffic analysis attacks involve communication pattern observations, traffic rate monitoring and time correlation of nodes' transmissions that are used to gain high level contextual information such as direction of traffic flow, traffic pattern and the source-destination location. Network-wide traffic rate monitoring attack [3][4] is done by eavesdropping and counting the number of transmitted/received packets around every node in the network, while network-wide time-correlation attacks [3][4] necessitate finding the general communication patterns or source-destination relationships.

A global view of network traffic patterns often reveal strong correlation, traffic contours or local maxima nodes that help global adversaries locate strategic nodes that are suspected sources and destinations [2][4]. Contextual privacy problem in wireless ad hoc networks, such as traffic direction inference, is mainly addressed by anti-traffic analysis techniques and context concealment methods that make it hard to trace back routes and localize source/destination. Link-by-link encryption scheme, commonly assumed in most research work, can only protect traffic content privacy [5]. Other methods, like padding to hide potential packet types through size and use of anonymous routing schemes cannot prevent traffic analysis attack completely. However, traffic cover with dummy packet insertion can obfuscate the real traffic patterns aspect, but has other implications on energy constraints, bandwidth utilization and overhead [6].

Strong privacy mechanism is crucial to protect source/destination nodes' location from target-oriented attacks and anonymize communication traffic to mitigate traffic analysis attacks. Source-destination pair privacy disclosure can be avoided by preventing an attacker to identify the true or exact starting node (i.e. source) and ending node (destination) of shortest path communication. In this work, we propose EXTROUT, Route Extrapolation, a simple and flexible privacy protocol to preserve source and destination privacy against a global powerful adversary. Source and destination nodes' privacy in large ad hoc wireless networks is achieved by camouflaging their transmission behavior among other identical broadcasting nodes. The two usual end nodes on a path (i.e. source and destination nodes) are concealed by extending the routing path by a random number of hops beyond destination node and source node, independently. EXTROUT also broadcasts dummy packets in an efficient way to make all nodes (including source and destination) along the extrapolated path appear indistinguishable. This will mask the origin and end of real packet flow. We compare EXTROUT with another commonly used privacy technique, called Fake Source-Destination pairs [10][11], and evaluate them against a global attacker model using privacy assessment metrics. There is trade-off between desired privacy strength and communication overhead cost. Hence, we also study how much privacy is achieved for different transmission overhead.

The rest of the paper is organized as follows: Section 2 describes related work on popular privacy techniques. Section 3 presents the network assumptions, attacker model and privacy metrics. The details of EXTROUT privacy protocol are explained in Section 4. Section 5 presents privacy strength analysis. Section 6 briefly explains fake source-destination pair set-up. Section 7 reports the simulation results and related discussions. Finally, Section 8 concludes the paper.

## II. RELATED WORK

Privacy mechanisms for ad hoc sensor networks have been studied previously [7][8] but a solution that achieves efficient and flexible privacy guarantees against global attacker remains elusive. Some privacy protocols proposed in the literature have considered the problem of location privacy either for the source node [9] [10] or the destination [2], independently, from a local eavesdropper, whilst other solutions deal with anti-traffic analysis techniques. It is more difficult implement source and destination privacy techniques in sparse traffic network, such as single source-destination case, against global eavesdropper.

Privacy protocols, such as [11][10][2][9][12][8], have focused on traffic pattern concealment methods for anonymous routing. Multipath traffic flows create some confusion about routes taken by packets to reach destination. The privacy performance when splitting packet flows in different or random ways spatially and temporally on multiple paths has been investigated in [12] using 'traffic entropy' measurement. But, a global attacker can still observe high traffic density near source region (as multipaths traffic flows diverge from source) and traffic accumulation when approaching destination region (as multipaths traffic flows converge towards destination). In baseline broadcast flooding [13], each node transmits the received packet to all its neighbor nodes in the network once. Flooding exhibits maximal coverage, redundancy and shortest path distance preservation. However, in real conditions, full scale flooding is energy expensive.

**Dummy Traffic:** The idea of dummy traffic in [3] is to properly insert dummy/fake packets in the real payload traffic stream so that the real traffic pattern is disguised. Source-destination communication characteristics are masked when dummy traffic adheres to a predefined pattern/behavior. With encrypted dummy packets as traffic padding, the adversary cannot even differentiate real packets from dummy packets. The delay experienced by real payload packets of a dummy protected flow depends on link bandwidth available. Thus, baseline dummy traffic generation increases communication overhead with extra energy consumption cost, while obfuscating node's real packet transmission occurrence. A constant rate dummy traffic over entire network will make all real packet transmissions undistinguishable, thus complicating rate monitoring (packet counting) attacks.

**Fake Source-Destination Pair:** The work in [11] initially proposed to create more replica sources (or fake sources) that input surplus dummy packets into the network traffic. Short-lived (temporary) fake sources provide better privacy strength against local attackers which cannot visualize the whole network topology and traffic flow. A long-lived (persistent) fake source node continuously generates dummy packets and gives a stronger privacy protection against global attacker with network-wide traffic knowledge. However, fake source-destination placement is decisive to create maximum uncertainty about the real source-destination location. For example, a fake source is normally positioned at the opposite direction with comparable route length as the real source-destination, and fake packet rate should also be comparable to that of the real source node [10]. The objective of fake source/destination nodes and dummy traffic method is to introduce unlinkability between source-destination traffic flow and communication direction that prevents contextual information disclosure.

In other related work [7], the authors proposed the use of two anonymous zones to preserve the anonymity of a source-destination pair, where a source or a destination is hidden anonymously among a group of other nodes. Flooding is confined in the local area of anonymous zones. In [14], the authors proposed SECLOUD privacy scheme, which also hides the source and destination in a group of nodes (called "clouds") that are indistinguishable in the presence of global traffic visualization attack. Mehta *et al.* [1] used a privacy scheme under global external attack model by hiding the real source among $k-1$ fake sources simulating the mobility pattern of real source in sensor network.

## III. MODELS AND ASSUMPTIONS

### A. Network Model

We assume a large ad hoc network model, typically consisting of hundreds of nodes that are distributed randomly or in a perturbed grid arrangement. The nodes are assumed quasi-static with sufficient battery power. Each node also has symmetric wireless links, i.e. has the same transmission and receiving range. Non-neighboring nodes communicate with each other via multi-hop links. At the medium access control (MAC) layer, we assume that each node runs IEEE 802.11 MAC interface in the "promiscuous" mode to receive all the MAC frames broadcasted in its neighborhood and all packet transmissions are also assumed to be locally broadcasted 1-hop distance range and not necessarily acknowledged. The existence of a key management protocol is implicit and pair-wise keys are distributed between nodes, for example using an efficient symmetric key protocol such as in [15][16]. All data and control packets transmitted have same size, are encrypted and no information is divulged to external attackers either through packet data or header content. Network nodes can distinguish different packet types (e.g. dummy packet from a real packet) by including a special tag which is itself encrypted in the packet; however the attacker cannot make this distinction. In [17], the authors used short-lived disposable MAC addresses to prevent the real node IDs from being revealed to adversaries. Finally, route discovery communications are assumed to be anonymous using any of the efficient anonymous routing protocols such as in [18][19]. Our assumptions are consistent with related works in the literature.

## B. Attacker model

In [1][12][20], the authors assumed global attacker with strong capabilities. A global attacker can eavesdrop every packet transmission in the network, trace the encrypted packets, and analyze network-wide traffic rate as well as correlate transmission times between nodes. Traffic contours, interconnection and intersection paths of online nodes can be inspected to divulge source and destination points. A number of inferences are derived from nodal packet count evidences and are used to estimate the probability of past packet movement itineraries. In this work, we do not consider attackers trying to disrupt the communication by jamming or route breaking or DoS attack. Hence, we also assume passive global attacker that performs rate monitoring and time correlation attacks for all traffic in the network. The attacker will visualize all transmitted/received packets in the network and determine traffic density on every link in the ad hoc network. However, the attacker cannot inject, modify or interrupt packet transmissions and can neither decrypt the content of captured packets. Since hop-by-hop packet re-encryption with different keys is applied, the attacker cannot trace the same packet on different links by examining the packet content. We focus on a global attacker with the endeavor to locate source and destination nodes in a wireless ad hoc network. This is similar to the idea in [12], where a global adversary is modeled as consisting of a number of colluding local adversaries that cooperate together and share information through a dedicated communication channel. An attacking network overlaid on top of the target network has also been considered in [1] as a realistic set up of the global adversary to eavesdrop the overall network communication.

## C. Privacy Metrics

The performance of EXTROUT is evaluated using anonymity and unlinkability metrics.

**Anonymity:** It means to obfuscate the real source/destination nodes in a group of indistinguishable nodes [21] [22] [23]. If the source node is hidden in a set of nodes $G_S$ and the destination is hidden in an independent group of $G_D$ nodes, the anonymity level of the source-destination pair is given as $\lambda = 1 - (1/G_S)(1/G_D)$.

**Unlinkability** [23] [24]**:** Privacy is maintained if packet transmissions are not linkable to any potential source or destination nodes. To test unlinkability, we employ traffic pattern visualization approach to identify hot spot traffic, traffic directionality/shape, local maxima nodes, traffic convergence and divergence. Perfect unlinkability between source-destination pair anonymous routing is comparable to a purely random guess, i.e. all nodes are in an indistinguishable state [24]. Uniformly distributed traffic in an ad hoc wireless network exhibits superior unlinkability performance under traffic analysis attacks.

## IV. EXTROUT PRIVACY PROTOCOL DESCRIPTION

In this section we describe the EXTROUT privacy protocol, which camouflage source-destination (*S-D*) nodes in ad hoc wireless networks. The idea is based on route extrapolation mechanism. Cover traffic, using dummy packets broadcasted by all nodes on the extrapolated route, masks the real packet transmission in the network at any given time. Traffic padding at link-level also conceals route discovery phase and initialization. For the functionality of EXTROUT protocol, we assume that all nodes have knowledge of the ad hoc network topology using any anonymous proactive routing protocol, such as [18][19], as mentioned before.

**EXTROUT** privacy protocol operates in three phases: (i) Initialization of *S-D* path, (ii) Route extrapolation source-side (to *Anchor_S*) and destination-side (to *Anchor_D*), and (iii) Dummy traffic generation.

### A. Phase 1- Initialization of S-D path

For setup and initialization, a route discovery process is initiated by source node *(S)* assembling a route request *(RQ)* packet and locally broadcasting it. Strong and efficient anonymous on demand routing protocols similar as reported in the literature [18] [25] can be used. The packet delivery path is set up after each node on the requested path replies and confirms the route to reach the destination. To prevent rate monitoring and time correlation attacks during initialization phase, route discovery packets can be concealed by periodically generating dummy traffic at a low rate (e.g. residual cover traffic) network-wide. Residual dummy traffic acts as "background noise" to conceal management and control packets transmissions during set up of source-destination path. For sensitive applications demanding maximum anonymity, this initialization overhead is worth the privacy gain. All nodes will transmit in broadcast mode at constant rate or probabilistic rate so that the attacker cannot link different transmission locations. An attacker can only infer position of transmissions using signal strength of packet and any existing localization algorithm.

### B. Phase 2 – S-D Route Extrapolation

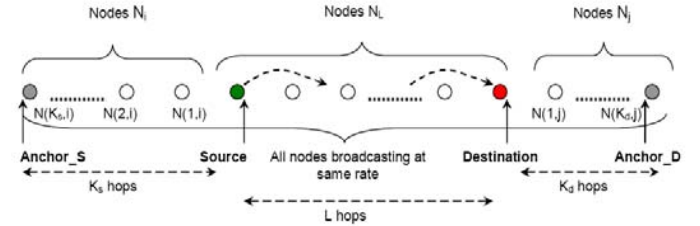

Figure 1. Route Extrapolation

Route extrapolation extends the shortest path, derived using Dijkstra algorithm**,** beyond the source (*S*) and destination (*D*) nodes on both sides. Fig. 1 shows the route extrapolation arrangement. Source node *S* first selects a *1st* hop neighbor node *i*, denoted as *N(1,i)*, excluding the node on shortest path *S-D* (of length *L*), that can reach node *D* via node *S* in *(L+1)* hops length. Next, node *S* selects a *2nd* hop neighbor node *N(2,i)* that can reach node *D* through *(L+2)* hops distance via node *S*. Subsequently, node *S* will select a list of nodes that extrapolates its shortest path *S-D* by linking successive nodes sequence *N(1, i), N(2, i), … , N(K_s, i)*. The furthest node *N(K_s, i)* from source *S*, called *Anchor_S* node, is $(L+K_s)$ hops away from the destination $D$ along the extended shortest path via the node $S$. The value of $K_s$ is set randomly by source $S$ so that the extrapolation distance from *Anchor_S* to node $S$ is unknown to the attacker.

Route extrapolation is applied in a similar manner by node $D$ on destination side. Destination node $D$ will select a list of nodes that extrapolates the shortest path $S-D$ to link successive nodes sequence from $N(1,j), N(2, j), …, N(K_d,j)$. The farthest node $N(K_d,j)$ from $D$, i.e. *Anchor_D* node, is $(L+K_d)$ hops away from source $S$ along the extended shortest path. The value of $K_d$ is set randomly by the destination $D$.

### C. Phase 3 - Dummy Traffic Generation

To obfuscate real packet transmission from eavesdropping attackers, dummy traffic (called "noise traffic") is generated at suitable rate, e.g. constant rate or probabilistic rate. All nodes on the extrapolated route from *Anchor_S* to *Anchor_D* will permanently broadcast background noise traffic by continuously generating dummy packets at a rate identical to the source $S$. As in [12][14], for EXTROUT protocol, all dummy (and real) packets have same size, are encrypted by different per-packet encryption keys at each node.

For a link $(x, y)$, node $x$ will broadcast a dummy packet, only next-hop node $y$ receiving it must identify and continue with another dummy packet broadcast, while other neighbor nodes of $x$ will ignore the dummy traffic received. For example in **Fig. 2,** with traffic flow from node '**a**' to node '**d**' at the rate of $A_1$ packets/second and another independent flow from '**b**' to '**d**' at the rate of $A_2$ packets/second, assuming $A_1 > A_2$. Cover traffic ($\varepsilon$) will be introduced on the link **a-b** by making node '**a**' broadcast dummy packets equal to $A_2$. As mentioned, per-link encryption mechanism would prevent an attacker from distinguishing dummy packets ($\varepsilon$) from real packets $(A_1$ or $A_2)$ and the attacker cannot trace back specific packets to determine which packet belongs to which source-destination traffic flow.

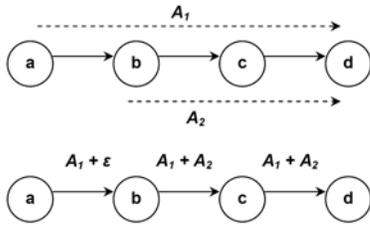

Figure 2. Dummy traffic

With the above set up, EXTROUT uses the shortest path (or an available direct path) to send packet to the destination $D$. Unusual high traffic nodes will not arise (except for nodes on path crossing) and there is uniform traffic density between *Anchor_S* and *Anchor_D*. Moreover, the edge/end nodes of the traffic are not necessarily the source or destination point. Extended path adds robustness against physical attacks on the source and destination node, e.g. in military ad hoc networks in battlefield. The attacker has to guess and destroy may be several nodes that are not clustered. Hence the physical attack to destroy the source or destination becomes expensive.

### D. EXTROUT with Path Duplication

Enhanced EXTROUT with path duplication method replicates the traffic flow between *Anchor_S* and *Anchor_D* using a number of disjoint routes between *Anchor_S* and *Anchor_D*. Let $P = \{P_1, P_2, …P_j\}$ be the set of $j$ disjoint paths from *Anchor_S* to *Anchor_D*, excluding the shortest path $L$. One or more disjoint paths $P_j$ are randomly selected and activated to emulate the traffic rate on path from *Anchor_S* to *Anchor_D*. Path $P_j$ behaves as a fake/duplicate path to increase probability of misleading attacker. EXTROUT with path duplication increases the anonymity level (at the cost of extra transmission overhead) by increasing the number of permutations for source-destination pairs on different paths perceptible to the attacker.

## V. PRIVACY PERFORMANCE ANALYSIS AND DISCUSSION

Keeping the extrapolated path longer and of random length improves anonymity, especially in single source-destination case when the network has low traffic load. Compared to other techniques in the literature, the EXTROUT preserves the privacy for uplink and downlink traffic of nodes.

### A. EXTROUT Privacy

The path length from source $S$ to destination $D$ is $L$ hops. $K_s$ and $K_d$ hops are the extended path segments on source side and destination side, respectively. Total path length from *Anchor_S* to *Anchor_D* is $(K_s + L + K_d)$ hops. Probability of selecting correct source/destination node is $\frac{1}{(K_s + L + K_d)}$. If EXTROUT with duplicated paths $(P_1, P_2,…P_n)$ have $n$ disjoint paths and assuming that the probability that one path being compromised/attacked is independent of others, then the probability of correctly selecting real traffic path and correct source-destination node pair will be $\left(\frac{1}{n+1}\right)\frac{1}{(K_s + L + K_d)}$.

EXTROUT protocol improves the *S-D* pair privacy (anonymity) statistically when more redundant (duplicate) disjoint paths are used between *Anchor_S* and *Anchor_D*. The protection of a path or a set of nodes from being compromised is another issue under our future work scope.

### B. EXTROUT Transmission Overhead

The transmission overhead factor (TOF) is considered as the ratio of total number of packets transmitted by nodes using privacy technique to the total transmissions without privacy technique using shortest path routing. TOF for baseline EXTROUT is given by $(K_s + L + K_d)/L$.

## VI. SHORTEST PATH WITH N-FAKE SOURCE-DESTINATION PAIRS

In fake source-destination privacy technique [1][11][12], a number of fake source-destination pairs $(S_f, D_f)$ are activated so that bogus communication route is established between them to persistently transmit dummy (fake) packets at the same rate as the real source $S$ whose location privacy we need to protect. The position of $(S_f, D_f)$ node pairs and the path length are important aspects to ensure effectiveness of this privacy enhancement scheme. Ideally, distance ($L_f$) between $S_f$ and $D_f$ must be comparable to distance ($L$) between $S$ and $D$ nodes. The anonymity set for source $S$ is the group of nodes at the extremities of the routing paths, i.e. $\{S, S_{f1}, S_{f2},...., S_{fn}\}$ and the destination anonymity set are the one-hop neighbors of the nodes $\{D, D_{f1}, D_{f2},...D_{fn}\}$, where $n$ is the number of fake ($S_f$, $D_f$) pairs. The source-destination node location anonymity level is $\{1 - (n+1)^{-1}\}$. Assuming real $S$-$D$ and all fake $S_f$-$D_f$ have same packet transmission rate, transmission overhead factor (TOF) with $n$ fake $S$-$D$ technique is given by $(L + \sum_{i=1}^{n} L_{fi})/L$. Fake source-destination pairs contributing to privacy enforcement must communicate as long as the real source has packets to send to its real destination.

## VII. SIMULATION AND EVALUATION

Our simulation based on NS-2 consists of a quasi-stationary ad hoc network of **400** nodes distributed in an area of **2000m × 2000m** with average node degree 7 (except for boundary nodes). Quasi-Unit disk graph (**Q-UDG**) connectivity model [26] is applied to simulate real world wireless network topologies. The (**x, y**) coordinates position of each node were randomly chosen in the ranges (**x-px$_i$ , x+px$_i$**) and (**y-py$_i$ , y+py$_i$**), respectively, where $p$ is a perturbation parameter and $x_i$ and $y_i$ are the spacing between the nodes in the $x$ and $y$ directions, respectively ( $p = 0.25$ and $x_i = y_i = i \times 100$ for $0 \leq i \leq 20$ were assigned for simulations). For more randomness in node distribution, we can increase the magnitude of perturbation parameter $p$, $(0 \leq p \leq 1)$. After nodes distribution, the **Q-UDG** connectivity model and transmission range are used to construct the network topology. A link exists between two nodes if the inter-nodal distance $d < \alpha R$, where $R$ is the transmission range of the node and $\alpha$ is the **Q-UDG** factor $(0 \leq \alpha \leq 1)$. In our simulations, we set $\alpha = 0.25$ and $R=145m$. For inter-nodal distance $d > R$, there is no link connectivity. But, for $\alpha R \leq d \leq R$, the link will exist with probability $(R-d)/(R-\alpha R)$. The source node (**S**) and destination node (**D**) are selected and a fixed number of real packets (e.g. 7000) are sent from **S** to **D** in a time window of **T** seconds. All simulations are repeated 20 times and results are averaged for each tested scenario. A matrix representation is used to indicate the approximate node positions in the network and the number of packets $u_i$ transmitted by each node $i$ (where $1 \leq i \leq 400$).

### A. Privacy with Fake Source-Destination

Fig. 3 shows the network matrix of number of packets transmitted by all nodes when using *1*-Fake **S-D** pair privacy method. The real source (**S**) and fake source (**S$_f$**), each sends 7000 packets (real packets by **S** and dummy packets by **S$_f$**) to the real destination (**D**) and fake destination (**D$_f$**), respectively, using the shortest path. By inspection, the traffic pattern reveals the location of the fake and real source/destination nodes which are at the ends of the single path. The global attacker can guess the real source (**D**) or real destination node (D) with probability ½, therefore anonymity level is ½, and the transmission overhead factor (TOF) is *2.08*.

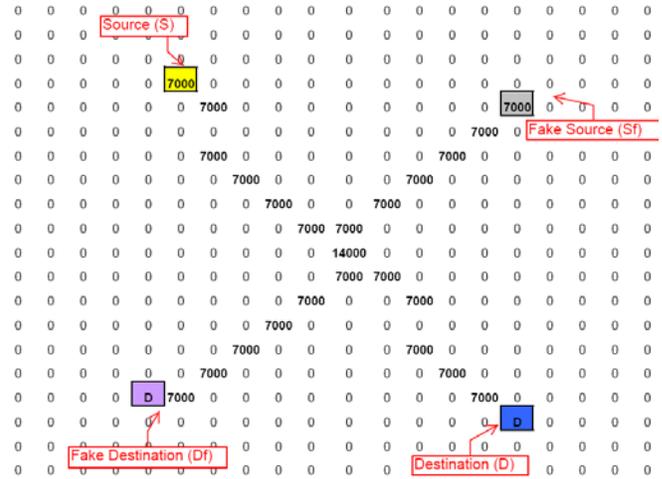

Figure 3. Single path with 1-Fake SD pair

### B. Privacy with EXTROUT Technique

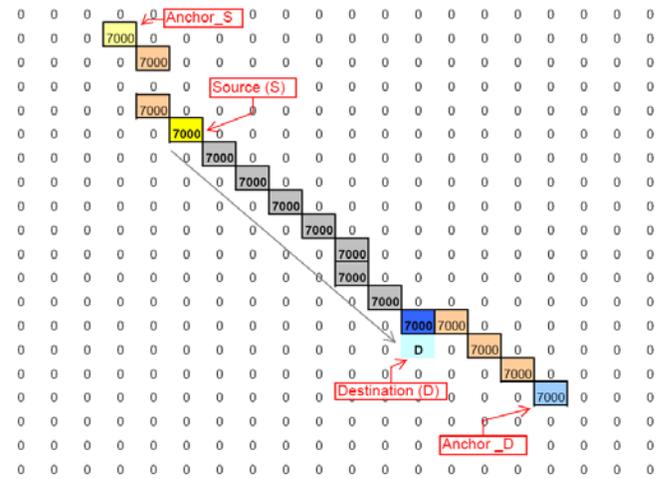

Figure 4. Baseline EXTROUT without duplicate path

Fig. 4 shows the active nodes with number of packets transmitted using EXTROUT for single source **S** and destination **D**, shortest path **L = 8** hops distance, $K_s = 3$ and $K_d = 4$. The attacker will guess location of the source/destination node out of the **15** indistinguishable nodes along the extended path. The anonymity level (privacy strength) for **S** or **D** is *[1-(1/15)], i.e 0.933*. The unlinkability is high because it is not clear for the attacker which node is the source (originator of packet transmission) and destination (receiver of real packets). The transmission overhead factor (TOF) is *1.875*.

In Fig. 5, we have EXTROUT with 1 path duplication for single **S-D** pair at **8** hops distance, $K_s = 3$ and $K_d = 4$. The attacker will guess location of the source/destination out of the two paths of **15** nodes length. The anonymity level is *0.967*. By inspection, the unlinkability of real **S-D** is higher and TOF = *3.75*. EXTROUT with **2** duplicated paths (diagram not shown for brevity) for the same single **S-D** pair at **8** hops distance, $K_s = 3$ and $K_d = 4$, gives anonymity level *0.978* and TOF is *5.625*.

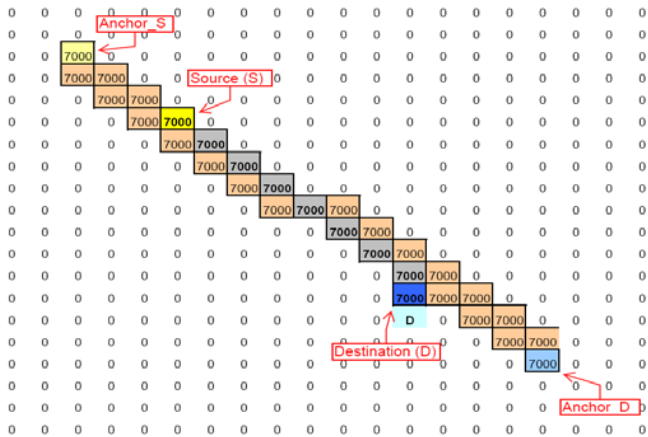

Figure 5. EXTROUT with 1 duplicate path

Fig. 6 shows EXTROUT with **5** duplicated paths **S-D** pair at **8** hops distance, $K_s = 3$ and $K_d = 4$. The **5** different paths have lengths **14, 15, 16, 16 and 19** hops between **Anchor_S** and **Anchor_D** (purple color cells). The attacker has to guess location of the source/destination out of the five paths, therefore anonymity level is **0.987** and TOF is **10**.

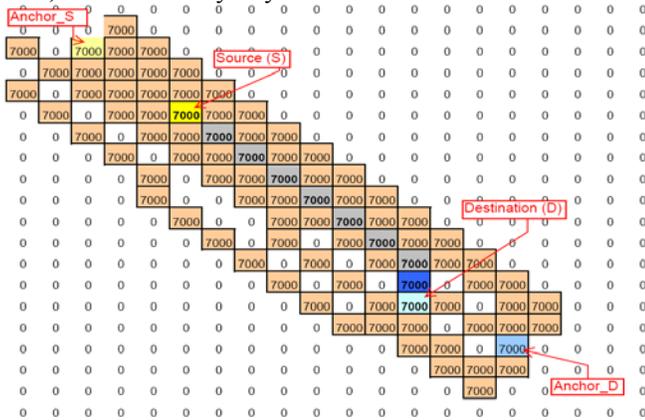

Figure 6. EXTROUT with 5 duplicate paths

### C. EXTROUT with Fake S-D Pair

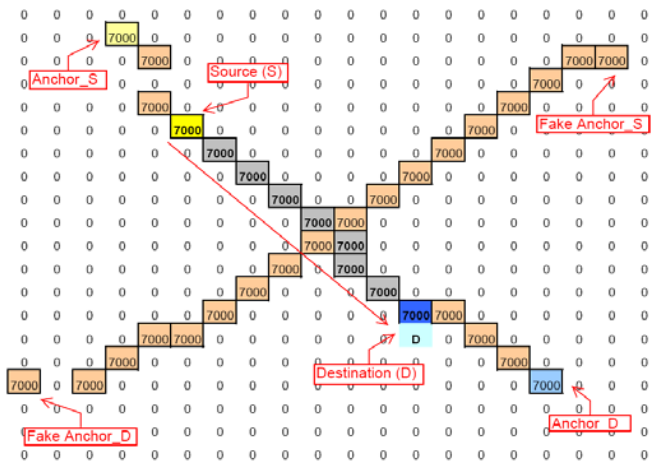

Figure 7. EXTROUT with 1-Fake SD pair

In Fig. 7, we show the result of EXTROUT with 1 fake S-D path added for single source-destination pair at **8** hops distance, $K_s = 3$ and $K_d = 4$. The fake extended path has length **17** hops between fake **Anchor_S** and fake **Anchor_D**. The attacker has to guess location of the source/destination by choosing one of the extended paths and then selecting the correct real **S-D** nodes pair, which gives anonymity level equal to **0.983**, i.e. **(**1/total length of both paths**)**. In this case, TOF is **4.25**. EXTROUT with fake path gives better performance (higher anonymity, lower transmission overhead) than EXTROUT with duplicate paths.

Fig. 8 shows the privacy level performance of EXTROUT for different source-destination distances (**L** hops). 100 randomly chosen source-destination pairs of different lengths (**3, 4, 5, …, 16** hops) are generated and EXTROUT privacy technique is applied. When the **S-D** path length is short (e.g. **3 to 5** hops), EXTROUT (with random value of $K_s=2$ & $K_d=2$) or (with random value of $K_s=2$ & $K_d=3$) has higher transmission overhead factor (TOF) and the corresponding privacy gain is lower. Using EXTROUT for source-destination nodes far apart (e.g. **> 10** hops) gives stronger privacy level **(> 0.96**) for the same transmission overhead cost.

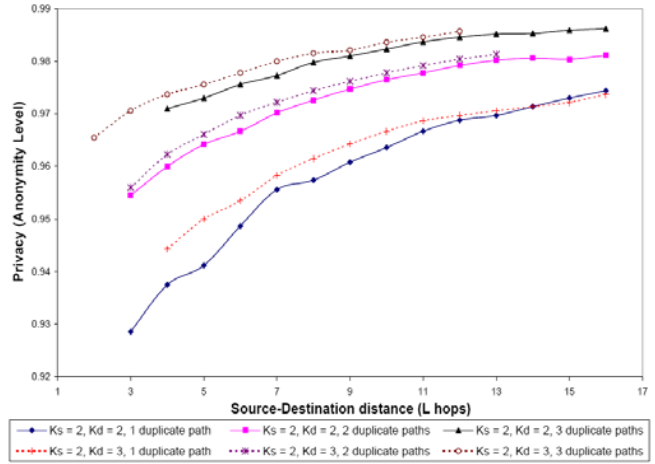

Figure 8. EXTROUT- Anonymity Performance

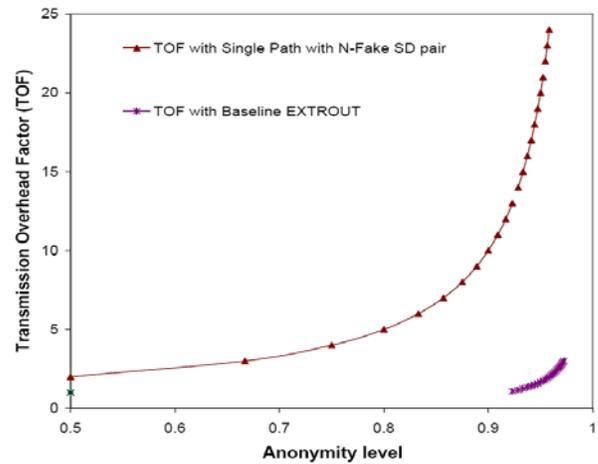

Figure 9. EXTROUT- Privacy Strength v/s TOF

EXTROUT mechanism is adaptive for small and large *S-D* separation in ad hoc wireless networks and for a given privacy level, the transmission overhead associated with it is lower than for single path with *N*-fake *S-D* pairs. Fig. 9 shows the relationship between anonymity level and TOF incurred by each privacy technique for single source-destination case at *L=12* hops.

## D. Results Discussion

EXTROUT privacy technique provides good performance tradeoff between anonymity level and transmission overhead. We summarize the results in Table I. Note $|P_n|$ is length of duplicate path and $|P_f|$ is the fake path length.

| Privacy Protocol | TOF | S-D Anonymity Level |
|---|---|---|
| Baseline EXTROUT | $(K_s+L+K_d)/L$ | $1-(K_s+L+K_d)^{-1}$ |
| EXTROUT with *n*-Path duplicate | $[(K_s+L+K_d)+\sum|P_n|]/L$ | $1-[(K_s+L+K_d)+\sum|P_n|]^{-1}$ |
| EXTROUT with *f*-fake | $[(K_s+L+K_d)+\sum|P_f|]/L$ | $1-[(K_s+L+K_d)+\sum|P_f|]^{-1}$ |
| N-Fake SD pairs | $(N+1)$ | $1-(N+1)^{-1}$ |

TABLE I. PRIVACY PROTOCOLS REVIEW

Increasing the length of extrapolated route, the anonymity of EXTROUT will increase but at the cost of extra transmission overhead. With EXTROUT privacy technique, the privacy is maintained as long as all nodes on extended path from *Anchor_S* to *Anchor_D* are alive (broadcasting), and the attacker can only guess the real source/destination node. With path duplication enhancement, EXTROUT will further camouflage real packet transmission route. EXTROUT has relatively moderate transmission overhead cost (which can vary) since the shortest *S-D* path is maintained (minimum delivery latency). The protocol provides flexibility of privacy level for source and destination nodes mutually.

## VIII. CONCLUSION

Providing maximum privacy in single source-destination case is a critical and challenging issue due to obvious traffic analysis attacks. In this work, an efficient and flexible privacy technique, EXTROUT, is presented to protect source-destination location privacy in the presence of a global adversary with network-wide traffic knowledge. Our analytical and simulation results show that EXTROUT provides better privacy (anonymity) to overhead ratio than simple fake source-destination privacy scheme in ad hoc wireless networks.

ACKNOWLEDGMENT

x

REFERENCES

[1] K. Mehta, D. Liu and M. Wright, "Location Privacy in Sensor Networks against a Glocal Eavesdropper" ICNP 2007.
[2] J. Deng, R. Han, and S. Mishra. "Decorrelating Wireless Sensor Network Traffic to Inhibit Traffic Analysis Attacks" Elsevier Pervasive and Mobile Computing Journal, Vol. 2(2), April 2006. pp. 159-186.
[3] M. Shao, Y. Yang, S. Zhu, G. Cao, "Towards Statistically Strong Source Anonymity for Sensor Networks" IEEE INFOCOM 2008, pp. 466-474.
[4] J. Deng, R. Han, and S. Mishra, "Countermeasures Against Traffic Analysis Attacks in Wireless Sensor Networks" in Proc. of IEEE/CerateNet Conf. on Security and Privacy in Communciation Networks (SecureComm) 2005.
[5] S. Capkun, JP Hubaux and M. Jakobsson, "Secure and privacy-Preserving Communication in Hybrid Ad Hoc Networks" EPLC-IC Technical Report no. IC/2004/10
[6] S. Jiang, NH Vaidya and W. Zhao, "Power Aware Traffic Cover Mode to Prevent Traffic Analysis in Wireless Ad Hoc Networks" IEEE INFOCOM 2001.
[7] X. Wu and E. Bertino, "An Analysis study on zone based Anonymous Communication in Mobile Ad Hoc Networks" EEE Trans. on Dependable and Secure Computing, Vol. 4, No. 4, 2007.
[8] Y. Jian, S. Chen, Z. Zhang and L. Zhang, "A Novel Scheme for Protecting Receiver's Location Privacy in Wireless Sensor Networks" IEEE Trans. on Wireless Comm.,Vol. 7, No. 10, Oct. 2008.
[9] Y. Xi, L. Schwiebert, and W. Shi, "Preserving source location privacy in monitoring-based wireless sensor networks," 2nd International orkshop on Security in Systems and Networks (SSN '06).
[10] P. Kamat, Y Zhang, W. Trappe, and C. Ozturk. "Enhancing source location privacy in sensor network routing". In Proc. of the 25th IEEE Int. Conf. on Distributed Comp. Systems (ICDCS), pp. 599-608, June 2005.
[11] C. Ozturk, Y. Zhang, and W. Trappe. "Source Location Privacy in Energy-constrained sensor network routing". In SASN '04: Proc. of ACM Workshop on Security of Ad Hoc and Sensor Networks, 2004.
[12] H. Choi, P. McDaniel, Thomas F. La Porta, "Privacy Preserving Communication in MANETs". IEEE SECON 2007
[13] C. Intanagonwiwat, R. Govindan, D. Estrin and J. Heidemann. "Directed Diffusion for Wireless Sensor Networking" IEEE/ACM Trans. on Networking, Vol 11, No. 1, 2003.
[14] R. Doomun, T. Hayajneh, P. Krishnamurthy, and D. Tipper, "SECLOUD: Source and Destination Seclusion Using Clouds for Wireless Ad Hoc Networks" In. Proc. of ISCC 2009.
[15] H. Chan, A. Perrig, and D. Song, "Random key predistribution schemes for sensor networks," in Proc. of IEEE SSP, 2003.
[16] W. Du, J. Deng, Y.S. Han, and P.K. Varshney, "A key predistribution scheme for sensor networks using deployment knowledge" IEEE Trans. On Dependable and Secure Comp., vol. 3, no. 1, pp. 62-77, 2006.
[17] M. Gruteser and D. Grunwald, "Enhancing location privacy in wireless LAN through disposable interface identifiers: a quantitative analysis" Mobile Network Applications, vol. 10(3), pp. 315-325, 2005.
[18] Y. Zhang, W. Liu, and W. Lou, "Anonymous communications in mobile ad hoc networks," in Proc. of IEEE INFOCOM, 2005.
[19] J. Kong, X. Hong, and M. Gerla, "An identity-free and on-demand routing scheme against anonymity threats in mobile ad hoc networks," IEEE Trans. on Mob. Comp., vol. 6(8), pp. 888–902, 2007.
[20] Y. Yang, M. Shao, S. Zhu, B. Urgaonkar, and G. Cao, "Towards Event Source Unobservability with Minimum Network Traffic in Sensor Networks" WiSec 2008.
[21] K. PN Puttaswamy, A. Sala, C. Wilson and B.Y. Zhao. "Protecting Anonymity in Dynamic Peer-to-Peer Networks" IEEE ICNP 2008.
[22] L. Sweeney "Achieving *k*-Anonymity Privacy Protection Using Generalization and Suppression" Int. Journal on Uncertainty, Fuzziness and Knowledge-based Systems, 10(5), 2002, pp. 571–588.
[23] D. Huang, "Traffic Analysis-based Unlinkability Measure for IEEE 802.11b-based Communication Systems", WiSe '06 Sept. 29, 2006. pp. 65-73.
[24] D. Huang, "Unlinkability Measure for IEEE 802.11 Based MANETs" IEEE Trans. on Wireless Communications, Vol. 7(3), 2008.
[25] J. Kong and X. Hong. ANODR: Anonymous On Demand Routing with Untraceable Routes for Mobile Ad-hoc Networks". Mobihoc'03, June 2003
[26] J. Chen, A. Jiang, Iyad A. Kanj, G. Xia, and F. Zhang, "Separability and Topology Control of Quasi Unit Disk Graphs" IEEE INFOCOM 2007.